\documentstyle[aps,manuscript]{revtex}

\begin{document}

\title{Synchronization in one-dimensional array of Josephson
       coupled thin layers}

\author{Dmitry A. Ryndyk}

\address{
Institute for Physics of Microstructures,
Russian Academy of Sciences \\
46 Ulyanov Str., GSP-105, Nizhny Novgorod 603600 Russia \\
E-mail: ryn@ipm.sci-nnov.ru}

\maketitle

\begin{abstract}
We obtain self-consistent macroscopic equations describing interlayer
Josephson effect and intralayer disequilibrium in one-dimensional array
of Josephson coupled layers.
We show that ``nonequilibrium coupling'' can lead to effective spatial and
time synchronization and formation of coherent dynamic resistive state
(collective Josephson effect) in
\mbox{Nb-AlO$_x$-Nb} stacked junctions and HTSC (intrinsic
Josephson effect). We propose it to be the origin of collective switching
phenomena observed \cite{a3} in PbBiSrCaCuO.
\end{abstract}


\null
It is well known that dc and ac Josephson effects take place in layered
superconductors when external current flows in direction perpendicular to the
layers. Two types of layered structures are currently experimentally
investigated: artificial Nb-AlO$_x$-Nb stacked junctions \cite{a1} and
natural layered high-T$_c$ superconductors \cite{a2}. This systems show
similar dynamic behavior \cite{a3} and can be considered on common
background. Electromagnetic theory of layered structures with thick enough
superconducting layers is developed in Ref.\ \cite{a4}, but in a case of thin
layers some other effects are to be considered, especially disequilibrium
which can be essential if layer thickness $d_0$ is smaller than
characteristic length of disequilibrium relaxation $l_E$ \cite{a5,a6,a7,a8}.
This criteria is obviously fulfilled for ``quantum'' structures with layers
of atomic thickness (HTSC). For ``classical'' artificial structures
\begin{equation}
l_E=\sqrt{\frac{2DT}{\pi \Delta^2_{\epsilon}}}\left(1+4\Delta^2_{\epsilon}
t^2_{\epsilon}\right)^{1/4}.
\end{equation}
Here $t_{\epsilon}$ is the inelastic electron-phonon scattering time,
$D=lv_F/3$ is the electron diffusion coefficient,
\mbox{$\Delta_{\epsilon}=(8\pi^2T_c^2\tau)/(7\zeta(3))$} is the energy gap,
$l$ is the free pass, $v_F$ is the Fermi velocity, \mbox{$\tau=(T_c-T)/T_c$}.
Thus $d_0<l_E$ can be fulfilled at least for $\tau\ll 1$. Last experiments
\cite{a3} manifest nonequilibrium effects in layered structures.

The origin of nonequilibrium Josephson effect is well established
\cite{a5,a6,a7,a8}.  First of all Josephson junction is a source of
disequilibrium in nonstationary state due to injection of quasiparticles.
This results in generation of nonzero covariant potential
$\Phi=\phi+(\hbar/2e)(\partial\theta/\partial t)$ in superconducting layers,
where $\phi$ is the electrostatic potential and $\theta$ is the phase of
superconducting condensate ($\Phi=0$ in equilibrium state). In nonequilibrium
regime an ordinary Josephson relation $(d\varphi/dt)=(2e/\hbar)V$ between
Josephson phase difference $\varphi=\theta_2-\theta_1$ and voltage
$V=\phi_1-\phi_2$ is violated. Instead we have
\begin{equation}
\label{NJR}
\frac{d\varphi}{dt}=\frac{2e}{\hbar}V+\frac{2e}{\hbar}(\Phi_2-\Phi_1).
\end{equation}
Thus disequilibrium modifies interlayer Josephson effect and self-consistent
description is necessary \cite{a7,a8}.

The simplest way to obtain self-consistent equations is Ginzburg-Landau
approach. In this paper we consider thin layer limit $d_0\ll l_E,\lambda_L$
and temperatures close to $T_c$, thus the discrete order parameter
$\Psi_n(x,y)$ for n-th layer can be used in a case of weak disequilibrium
($|\Psi_n|\approx const$). Static equations of Ginzburg- Landau type for
$\Psi_n$ are obtained by Lowrence and Doniach \cite{a9} (see also Ref.\
\cite{a10,a11,a12,a13}).  Here we use dynamic generalization of their model in
one-dimensional case \mbox{$\Psi_n(x,y,t)=\Psi_n(t)$} (three dimensional
variant can be derived straightforward). Thus we consider the structure with
spatial dimensions less than characteristic magnetic length and without
external magnetic field, so that Josephson junctions are effectively
zero-dimensional.
One-dimensional state is also possible in larger systems but stability
analysis is necessary.
Interlayer interaction is assumed to
be small (only two neighbor layers interact).

Model time-dependent Ginzburg-Landau equation of Lowrence-Doniach form may be
obtained from relaxation principle
\begin{equation}
\label{relax}
\gamma(\hbar\frac{\partial}{\partial t} +2ie\phi_n)\Psi_n=-\frac{\delta
F\{\Psi_n\}} {\delta\Psi_n^*}.
\end{equation}
Here we introduce electrostatic potential of n-th layer $\phi_n$. Free energy
$F\{\Psi_n\}$ is (we can choose the gauge {\bf A}=0 in one-dimensional case)
\begin{equation}
\label{free}
F\{\Psi_n\}=d\sum_n\int dxdy\left[-\alpha\tau
|\Psi_n|^2+\frac{b}{2}|\Psi_n|^4 +\alpha r|\Psi_{n+1}-\Psi_n|^2\right],
\end{equation}
where $\alpha$ and $b$ are GL constants, $d$ is in general some dimensional
parameter depending on $\alpha$ and $b$ definition and coinciding as a rule
with spatial period of a structure ($d\approx d_0$ for ``classical'' S-I-S
structures and interlayer distance for ``quantum'').  $r$ is dimensionless
parameter of interlayer interaction.  The microscopic expression for $r$ is
rather distinct in ``quantum'' and ``classical'' cases.  But from
phenomenological point of view it is more appropriate to express $r$ from
equation (\ref{J}) used below
\begin{equation} r=\frac{8\pi
e\lambda_{0\parallel}^2\xi_{0\parallel}^2J_0}{\hbar c^2 d},
\end{equation}
where $\lambda_{0\parallel}$ and $\xi_{0\parallel}$ are parallel to the
layers London and coherence lengths at zero temperature and
$J_0=(4e\alpha^2dr)/(\hbar b)$ is Josephson critical current density at zero
temperature.  In both cases if $r\ll 1$ then only two neighbor layers
interact.

 From (\ref{relax}),(\ref{free}) we obtain equation for dimensionless order
parameter $\psi_n=\Psi_n/\sqrt{\alpha\tau/b}$
\begin{equation}
\label{LD}
-t_{GL}\left(\frac{d}{dt}+i\frac{2e}{\hbar}\phi_n\right)\psi_n+
(1-|\psi_n|^2)\psi_n-\epsilon(2\psi_n-\psi_{n+1}-\psi_{n-1})=0.
\end{equation}
$\epsilon=r/\tau$ is coupling parameter, it characterizes a ratio of
Josephson energy to energy of superconducting condensate.
In the weak coupling limit $\epsilon\ll 1$, $|\psi_n|\approx 1$ introducing
$\psi_n=e^{\displaystyle i\theta_n}$,
$\displaystyle\Phi_n=\phi_n+\frac{\hbar}{2e}\frac{d\theta_n}{dt}$,
\mbox{$\varphi_n=\theta_{n+1}-\theta_n$} we obtain from (\ref{LD}) the
expression for covariant potential $\Phi_n$
\begin{equation}
\label{dyn1}
t_{GL}\frac{2e}{\hbar}\Phi_n=
\epsilon(\sin\varphi_n-\sin\varphi_{n-1}),
\end{equation}
where $t_{GL}=t_0/\tau$,
\mbox{$t_0=(\gamma\hbar)/(\alpha d)$} is characteristic relaxation time and
can be found to be (at $|\psi|=const$) \cite{a15}
\begin{equation}
t_0=\frac{\pi\hbar}{8T_c}\frac{1}{\sqrt{1+\Gamma_0^2\tau}},
\end{equation}
where $\Gamma_0=\sqrt{2/7\zeta(3)}\pi\hbar^{-1} T_c t_{\epsilon}$.  Note,
that this expression follows from time-dependent Ginzburg-Landau equation
derived in Ref.\ \cite{a15} from microscopic theory in the so-called local
equilibrium approximation, and this equation coincides with our model
equation in weak disequilibrium and ``discrete'' case.

Expression for
superconducting current can be obtained from variation principle (in 3D form)
and full current between the n-th and n+1-th layers is sum of supercurrent,
normal current taken in simplest ohmic form and capacity current due to
polarization of interlayer space
\begin{equation}
\label{J}
J_{n,n+1}=-\frac{i}{2}J_c(\psi_{n+1}\psi_n^*-\psi_{n+1}^*\psi_n)-
\frac{(\phi_{n+1}-\phi_n)}{R}-C\frac{d}{dt}
(\phi_{n+1}-\phi_n)=const,
\end{equation}
here $J_c=J_0\tau$, $R_{n,n+1}$, $C$ are the dimensional parameters. We
neglect self charge of superconducting layers (which is in general small due
to quasineutrality) and thus one-dimensional current is conserved (possible
peculiarities in dynamics of ``quantum'' structures with $d_0$ compared with
Debye length will be considered in other paper, see also Ref.\ \cite{a14}).
We introduce random resistivity $R_{n,n+1}=R_0/\delta_{n,n+1}$ to
account the effects of nonidentical junction parameters in simplest way.
Using (\ref{NJR}) which is direct consequence of $\Phi_n$ definition we obtain
from (\ref{J})
\begin{equation}
\label{dyn2}
\omega_p^{-2}\frac{d^2\varphi_n}{dt^2}+
\omega_c^{-1}\delta_n\frac{d\varphi_n}{dt}+\sin\varphi_n
-\omega_c^{-1}\frac{2e}{\hbar}(\Phi_{n+1}-\Phi_n)-\omega_p^{-2}
\frac{2e}{\hbar}\frac{d}{dt}(\Phi_{n+1}-\Phi_n)=j,
\end{equation}
where $\omega_p^2=(2eJ_c)/(\hbar C)$,
$\omega_c=(2eR_0J_c)/\hbar$, $j=J/J_c$.
As a result we have a set of coupled equations (\ref{dyn1}), (\ref{dyn2})
describing Josephson effect and disequilibrium in self-consistent manner.
This equations are in fact more general than
underlying time-dependent Lowrence-Doniach equation (\ref{LD}). Eq.
(\ref{dyn1}) demonstrates well known result $\Phi\propto div {\bf j}_s$ and
can be derived at any temperature. Eq. (\ref{dyn2}) is a sequence of basic
nonequilibrium Josephson relation (\ref{NJR}) and expression for current
between layers (\ref{J}). Thus we think that proposed in this paper
macroscopic model is a good basis for describing nonequilibrium Josephson
effect in layered superconductors.

Equations (\ref{dyn1}), (\ref{dyn2}) are equivalent to a set of coupled
Josephson equations for $\varphi_n$ with additional interaction terms
\begin{eqnarray}
&\displaystyle\beta\frac{d^2\varphi_n}{dt'^2}+\delta_n\frac{d\varphi_n}{dt'}+
\sin\varphi_n+f\{{\varphi}\}=j,& \\
&\displaystyle f\{\varphi\}=\eta(2\sin\varphi_n-\sin\varphi_{n+1}-
\sin\varphi_{n-1}) +
\beta\eta(2\cos\varphi_n\frac{d\varphi_n}{dt'}-\cos\varphi_{n+1}\frac
{d\varphi_{n+1}}{dt'}-\cos\varphi_{n-1}\frac{d\varphi_{n-1}}{dt'}).&
\end{eqnarray}
$\beta=\omega_c^2/\omega_p^2$, $t'=\omega_ct$,
$\eta$ is the novel parameter of nonequilibrium coupling
\begin{equation}
\eta=\frac{\epsilon}{\omega_c t_{GL}}=\frac{r}{\omega_{c0}t_0\tau}=
\frac{8T_c r\sqrt{1+\Gamma_0^2\tau}}{\pi\hbar\omega_{c0}\tau}.
\end{equation}
Let us estimate $\eta$ for some real structures. First of all
$\Gamma_0^2$ is usually large as $(T_D/T)^4$, $T_D$ is Debye temperature
$$t_{\epsilon}\sim\frac{\hbar}{T}\left(\frac{T_D}{T}\right)^2\ \ \
\rightarrow\ \ \ \Gamma_0\sim\left(\frac{T_D}{T_c}\right)^2\sim
\left\{\begin{array}{lll}
10^3 & Low\ T_c\ SC& \Gamma_0^2\sim 10^6 !\\
10   & HTSC        & \Gamma_0^2\sim 100 \end{array}\right.$$
And for $\eta$ we obtain
\begin{equation}
\eta=\frac{8T_c r\Gamma_0}{\pi\hbar\omega_{c0}\sqrt{\tau}}.
\end{equation}
We find that $\eta\sqrt{\tau}\sim 1$ for HTSC at $r\sim 10^{-3}$,
$\omega_{c0}\sim 10^{12}$rad/sec and for \mbox{$Nb-AlO_x-Nb$} structures at
$d_0\sim 100 \AA$, $R_NS\sim 10^{-8}$ $\Omega$ cm$^2$ ($R_N$ - resistivity,
$S$ - area of Josephson junction).
Thus we see that weak disequilibrium can lead to strong coupling between
interlayer Josephson junctions and to collective dynamic behavior in a case
of weak coupling between layers ($\epsilon\ll 1$).
In a case of nonidentical junctions this lead to synchronization.
Simulations show that
at $\eta > 1$ and $\beta\eta < 1$ coherent dynamic state with
$\dot{\varphi}_n(t)=\dot{\varphi}(t)$ is
single and stable. Full investigation of this problem is a subject of
future work.

Presented theory is able to explain recent experiments \cite{a3} on
resistivity (at current perpendicular to the layers) of BiSrCaCuO and
PbBiSrCaCuO superconductors. In Bi-compound many hysteresises on voltage --
current characteristic are observed due to one by one switching of interlayer
junctions and in PbBi-compound all junctions are switched to resistive state
simultaneously. The estimation of parameter $\eta$ for this samples gives
$\eta\sim 0.1\div 1$ for the first of them and $\eta\sim 0.5\div 5$ for
the second one. Thus collective switching may be induced by nonequilibrium
coupling.

This work was supported by the Russian Foundation for Basic Research, grant
No. 93-02-14718 and by a fellowship of INTAS Grant No. 93-2492 and is carried
out within the research program of International Center for Fundamental
Physics in Moscow. The author thanks Prof. A. A. Andronov, Dr. V. V. Kurin
and Dr. A. S. Mel'nikov for useful and stimulated discussions.


\begin{references}
\bibitem{a1} H. Kohlstedt, G. Hallmanns, I. P. Nevirkovets, D. Guggi, and
C. Heiden, IEEE Trans. Appl. Supercond. {\bf AS-3}, 2117 (1993);
I. P. Nevirkovets, H. Kohlstedt, G. Hallmanns, and C. Heiden,
Supercond. Sci. Technol. {\bf 6}, 146 (1993); A. V. Ustinov, H. Kohlstedt,
M. Cirillo, N. F. Pedersen, G. Hallmanns, and C. Heiden, Phys. Rev. B
{\bf 48}, 10614 (1993).
\bibitem{a2} R. Kleiner and P. M\"{u}ller, Phus. Rev. B {\bf 49}, 1327 (1994).
\bibitem{a3} R. Kleiner, P. Muller, H. Kohlstedt, N. Pedersen,
and S. Sakai, Phys. Rev. B {\bf 50}, 3942 (1994)
\bibitem{a4} S. Sakai, P. Bodin, and N. F. Pedersen, J. Appl. Phys. {\bf 73},
2411 (1993).
\bibitem{a5} {\em Nonequilibrium Superconductivity}, edited by D. N.
Langenberg and A. I. Larkin, Modern Problems in Condensed Matter Science
Vol. 12 (North-Holland, Amsterdam, 1986).
\bibitem{a6} A. M. Gulyan and G. F. Zharkov, {\em Superconductors in
external fields (nonequilibrium phenomena)} (Moscow, 1990, in russian).
\bibitem{a7} S. N. Artemenko and A. F. Volkov, Usp. Fiz. Nauk {\bf 128}, 3
(1979).
\bibitem{a8} A. M. Gulian and G. F. Zharkov, Zh. Eksp. Teor. Fiz
{\bf 89}, 156 (1985).
\bibitem{a9} W. E. Lowrence and S. Doniach, in Proceedings of the 12th Int.
Conf. on Low-Temperature Physics, Kyoto 1970, p. 361.
\bibitem{a10} L. N. Bulaevskii, Usp. Fiz. Nauk {\bf 116}, 449 (1975)
[Sov. Phys. Uspekhi {\bf 18}, 193 (1975)].
\bibitem{a11} L. N. Bulaevskii, Zh. Eksp. Teor. Fiz. {\bf 64}, 2241 (1973)
[Sov. Phys. JETP {\bf 37}, 1133 (1973)].
\bibitem{a12} L. N. Bulaevskii, Adv. Phys. {\bf 37}, 443 (1988);
Int. J. Mod. Phys. B {\bf 4}, 1849 (1990).
\bibitem{a13} R. A. Klemm, M. R. Beasley, and A. Luther, J. Low Temp. Phys.
{\bf 16}, 607 (1974)
\bibitem{a14} L. N. Bulaevskii, M. Zamora, D. Baeriswyl, H. Beck, and
J. R. Clem, Phys. Rev. B {\bf 50}, 12831 (1994)
\bibitem{a15} R. J. Watts-Tobin and L. Kramer, Phys. Rev. Lett.
{\bf 40}, 1041 (1978).
\end{references}
\end{document}